\documentclass[12pt,a4paper]{article}
\usepackage{amsmath}
\usepackage{amssymb}
\usepackage{amsfonts}
\usepackage{epsfig}
\title{Percolation and lack of self-averaging in a frustrated evolutionary
model}
\author{Andrea De Martino\footnote{E-mail address:
dmart@physig.phys.uniroma1.it.}~
and Andrea Giansanti\footnote{E-mail address: giansanti@roma1.infn.it.}\\
{\normalsize \emph{Dipartimento di Fisica, Universit\`a di Roma ``La Sapienza'',}}\\
{\normalsize \emph{Piazzale A. Moro 2, 00185 Roma, Italy}}}
\date{}
\begin{document}
\maketitle
\noindent
{\small \textbf{Abstract -}}
{\small We present a stochastic evolutionary model obtained through a
perturbation
of Kauffman's maximally rugged model, which is recovered as a special
case. Our main results are: (i) existence
of a percolation-like phase transition in the finite phase space case;
(ii) existence of non self-averaging effects in the thermodynamic
limit. Lack of self-averaging emerges from a fragmentation of the space
of all possible evolutions, analogous to that of a geometrically broken
object. Thus the model turns out to be exactly solvable in the
thermodynamic limit.}\\
\mbox{}
\newline
\\
\section{Introduction}
We present here the analytic study of a model of an abstract behaviour
with frustrated rationality. The model, despite or because of its
ingenuity, has revealed interesting statistical properties, such as a
percolative phase transition in the finite dimensional case and
non self-averaging effects in the thermodynamic
limit. Our starting point is Kauffman's well known $NK$ model of
biological evolution (see Ref. \cite{NK}) in its maximally rugged
version ($K=N-1$),
whose properties have been extensively investigated.
Yet in this paper it serves as a metaphoric abstract model for
the behaviour of a fully rational adaptive walker who moves in its phase
space in search for an optimal configuration. We decided to perturb its
stringent rationality by introducing in the evolutionary rule a probability
$p$, as a measure of a certain degree of insanity (or frustration or
disorder). For $p=1$ we recover the original model, whereas for $p=0$ we have
a random walker in configuration space. The introduction of $p$ is fatal
for the  adaptive behaviour, but leads to a percolation-like phase
transition, that separates a phase characterized by finite walks to
optima from one in which the probability of an interminable
walk is non zero. We show that in a large configuration space a small
perturbation is sufficient to get the percolation threshold.
The thermodynamic limit is obtained by letting the cardinality
of the phase space go to infinity.
We argue that this leads to an infinite number of different
possible evolutions.
Nevertheless, in this limit we show that the probability
$Y$ that two walkers undergo ``similar'' (in a sense that will become clear
later) evolutions has non zero average and a finite variance,
that is it lacks of self-averaging.
This property will result from
a fragmentation of the space of all possible evolutions analogous to that
of a geometrically broken object (Ref. \cite{adm}).

Evolutionary models have become quite familiar to theoretical physicists,
and many of them have been carefully examined
(see Refs. \cite{peliti,higgs} for a review). This for two
main reasons. (i) Species undergoing biological evolution are dynamical
systems, in the sense that their configuration varies with time according to
some modelizable dynamical law. The dynamics draws a trajectory in
the system's phase space, that is the set of all possible configurations.
(ii) Biological evolution is a complex phenomenon, since one must assume that
each step of it derives from and is influenced by the concurrence of different
factors, which may be altogether taken into account as a number of
random variables that give the system's trajectory unpredictability and
stochasticity. Hence, many ideas taken from the theory of disordered
systems, the main of which is that of \emph{landscape}, may be fruitfully
applied for the construction and the study of these seemingly different types
of models (see Ref. \cite{physicad} for a detailed overview).

One assumes that an evolving species (the system) 
may be found in any of a number of configurations, representing its
genome. This is taken for simplicity to be a finite set of spin
variables $S_i$ ($S_i=\pm 1$, $1\leq i\leq N$). The phase space
$\Gamma$ is the set $\{+1,-1\}^N$ of all genomes. The metric in $\Gamma$
is typically the Hamming distance $d_H$.
Feature (i) is recovered by giving some evolutionary algorithm
$F$ such that
\begin{equation}
\mathcal{C}_{t+1}=F(\mathcal{C}_{t}),
\end{equation}
where $\mathcal{C}_t$ is the system's configuration at time $t$ and time
is a positive integer or zero. Feature (ii) is introduced through the
concept of landscape. For our purposes, a landscape is a pair
$(\Gamma,\phi)$, $\Gamma$ being the system's phase space $\phi$ being a
real valued function $\phi:\Gamma\rightarrow\mathbb{R}$ called
\emph{fitness}, defined for all $\mathcal{C}\in\Gamma$.
The idea underlying biological evolutionary models is that $F$ lets the
system evolve through configurations of growing fitness in search for an
optimal one. This optimization procedure is usually not global, 
that is the system does not seek for the fittest configuration in
$\Gamma$; optimal configurations are considered those $x\in\Gamma$ such that
$\phi(x)>\phi(y)$, for all $y\in\Gamma$ such that $d_H(x,y)=1$. These are
called ``local optima''.

Of course, the complexity arises from the difficulty in
finding the local optima, or, if one wants, from the specific form
of $\phi$, which may eventually depend on $t$. The more rugged the landscape,
namely the higher the number of maxima and minima of $\phi$, the more complex
the dynamics.

In Kauffman's original idea the fitness of each configuration resulted from
epistatic interactions between $K$ of its $N$ genes. An increase of $K$
implied an increase of the number of local fitness optima. This
way of tuning the landscape's complexity is equivalent to the following,
which may sound more familiar to physicists
(see Ref. \cite{stein} for an overview of the contact points between
spin glass physics and biology).
One assumes that the fitness of a configuration $\mathcal{C}$ is given by
a $K$-spin type of hamiltonian,
\begin{equation} \label{fitness}
\phi(\mathcal{C})=\sum_{i_{1},\ldots,i_{K}=1}^{N}J_{i_{1},\ldots,i_{K}}
S_{i_{1}}\cdots S_{i_{K}},
\end{equation}
where $\mathcal{C}=\{S_1,\ldots,S_N\}$ and $J_{i_1,\ldots,i_K}$ are
gaussian random variables, with $K\leq N$.
It is possible to show (see Refs. \cite{rem1,rem2}
for details) that as the parameter $K$ varies from
$1$ to $N$ the landscape's ruggedness grows accordingly, since correlations
between the fitness' values of neighbouring configurations
(configurations $x$ and $y$ such that $d_H(x,y)=1$) decrease. Therefore
for large $K$, that implies large $N$, one finds that the probability
$P(\phi_1,\phi_2)$ that two configurations $\mathcal{C}_1$ and
$\mathcal{C}_2$ have fitnesses $\phi_1\equiv\phi(\mathcal{C}_1)$ and
$\phi_2\equiv\phi(\mathcal{C}_2)$ respectively factorizes:
\begin{equation}
P(\phi_1,\phi_2)\simeq P(\phi_1)P(\phi_2).
\end{equation}
For all practical purposes $\phi$ behaves thus as a random variable.
This is Kauffman's maximally rugged landscape, which is equivalent to
Derrida's random energy model (again Refs. \cite{rem1,rem2}).

This paper is organized as follows.
In Section 2 we give an account of Kauffman's maximally rugged model,
with its main statistical properties. Though much of the material of
Section 2 are well-established results, we added them to this paper both to
make it self-consistent and to emphasize how the perturbation acts on
the system. Section 3 contains the definition of the perturbed model and its
analytic study in the finite phase space case. In Section 4 we study the
thermodynamic limit. In the final section we make some comments on our
results and formulate the conclusions.

\section{Kauffman's maximally rugged model}
Kauffman's maximally rugged model is defined as follows: the system may
take on any of the $2^N$ configurations of the phase space
$\Gamma=\{+1,-1\}^N$, and large $N$ is assumed. The fitness $\phi$ is a
quenched random variable whose probability density is, say, $p(\phi)$.
The dynamics $F$ is then defined as a zero temperature Monte Carlo algorithm:
\begin{enumerate}
\item at time $t\geq 0$ the system is in configuration $\mathcal{C}_t=\{S_1,
\ldots,S_N\}$ with fitness $\phi(\mathcal{C}_t)$;
\item a spin $S_i$ of $\mathcal{C}_t$ is chosen at random and its sign is
changed, thus obtaining a configuration $\mathcal{C'}$ that differs from
$\mathcal{C}_t$ by just the $i$-th spin ($1\leq i\leq N$);
\item if $\phi(\mathcal{C'})>\phi(\mathcal{C}_t)$ then $\mathcal{C}_{t+1}=
\mathcal{C'}$; otherwise $\mathcal{C}_{t+1}=\mathcal{C}_t$ and return
to (ii).
\end{enumerate}

In a rough biological interpretation, this models a situation in which a
species evolves increasing its fitness by random point mutations.
Trajectories come to an end when the system is in a local fitness
maximum, because it cannot find any fitter neighbour.
$F$ is a stochastic dynamics that takes the system
to such optima passing through configurations of increasing fitness that
are just one spin different from one another. The trajectories are
usually called \emph{adaptive walks}, and their length is strictly
related to the local properties of the fitness landscape. These
have been analytically studied (see Refs. \cite{NK} and
\cite{weinberger}),
revealing a generous structure of very numerous maxima, as we shall soon
recall.

Before coming to that, we would like to stress that in what follows we
shall consider two types of averages. The first one, which
we shall call a ``quenched'' average, will be denoted by a bar
($\overline{\cdots}$) and indicates averages over all possible
fitness realizations.
Suppose to be given a certain quantity $q$ (for instance, the number of
fitness maxima) that may take on different
values in different realizations of $\phi$. The average of $q$ over
all possible fitness samplings will be written $\overline{q}$.
This notation is slightly unusual since generally this type of average is
denoted by brackets. Instead, the second
one will be denoted here by brackets ($\langle\ldots\rangle$) and will
define averages over many different evolutions. For instance, we shall
deal with the average length $\langle\ell\rangle$ of an adaptive walk.
This could be written $\langle\ell\rangle=\int\ell dQ(\ell)$, where
$Q(\ell)$ is the probability that an adaptive walk consists of $\ell$
steps. In principle it may be difficult to obtain analytical information
about the probabilities $Q(x)$ ($x=$ length, duration, \ldots).
This average may nevertheless be estimated as follows: one
fixes the landscape and averages the lengths of many walks with the same
starting point, which by assumption will be the least fit
configuration in $\Gamma$. The ensemble in which averages are calculated
are thus that of all possible landscapes on $\Gamma$ for $\overline{\cdots}$
and that of all possible evolutions (for example, adaptive walks) for
$\langle \ldots\rangle$.

We begin by proving that the average number of local optima increases
exponentially with $N$.
\begin{description}
\item[Result 1.]Let $f(N)$ denote the fraction of local fitness optima in
$\Gamma$ in a given fitness realization (landscape). We have
\begin{equation}
\overline{f(N)}=\frac{1}{N+1}.
\end{equation}
\end{description}
The proof is straightforward: let $y(\phi)$ denote the probability that
a given configuration has a lower fitness than $\phi$, namely
$y(\phi)=\int_{-\infty}^{\phi}p(\phi')d\phi'$. Since for a local optimum
of fitness $\phi$ the $N$ neighbouring configurations must have lower fitness,
we have that $\overline{f(N)}$ is the average of $y^N$ over all possible
choices of $y$. The probability density $q(y)$ of $y$ is uniform, hence
\begin{equation}
\overline{f(N)}=\int_0^1 y^N dy=\frac{1}{N+1}.
\end{equation}

On the average there are thus $2^{N}/(N+1)$ local optima, so that their
number grows exponentially with $N$. Now consider
making $\ell$ steps of an adaptive walk starting from the least fit
configuration in $\Gamma$. One finds that on the average the probability
to take a further step, namely the fraction of fitter neighbours, is halved
each time a step is taken.
\begin{description}
\item[Result 2.] Let $F(\ell)$ denote the fraction of fitter
neighbours after $\ell$ steps. We have
\begin{equation}
\langle F(\ell)\rangle=2^{-\ell}.
\end{equation}
\end{description}
Indeed, an adaptive walk can be seen as a sequence of increasing
but independent values
of $\phi$. If we consider $y(\phi)$ instead of $\phi$, an adaptive walk
becomes a
sequence of increasing values of $y$, which is, as said above, a
random variable with
uniform probability density on the $[0,1]$ interval. For one walk of
$\ell$ steps, namely for one increasing sequence of $\ell$ independent values
$y_1$,\ldots,$y_\ell$ of $y$, we can write the probability to find
an $(\ell+1)$-th value of $y$ greater than all of the previous $\ell$ as
\begin{equation}
F(\ell)=P(y_2>y_1)P(y_3>y_2)\cdots P(y_{\ell+1}>y_\ell) ,
\end{equation}
where $P(y_n>y_m)$ denotes the probability of sampling a value $y_m$ of $y$
greater than $y_n$.
$\langle F(\ell)\rangle$ may be obtained by averaging over all possible
samplings of $y_1,\ldots,y_\ell$:
\begin{equation}
\langle F(\ell)\rangle=
\langle P(y_2>y_1)P(y_3>y_2)\cdots P(y_{\ell+1}>y_\ell)\rangle .
\end{equation}
Clearly, $P(y_{n+1}>y_n)=1-\int_0^{y_n}q(y)dy=1-y_n$,
because $q(y)$ is uniform. Hence
\begin{equation}
\langle F(\ell)\rangle=\langle(1-y_1)\cdots(1-y_\ell)\rangle.
\end{equation}
The statistical independence of $y_1,\ldots,y_\ell$ implies that
\begin{equation}
\langle F(\ell)\rangle=
\langle(1-y_1)\rangle\cdots\langle(1-y_\ell)\rangle=
\langle(1-y)\rangle^\ell.
\end{equation}
Now it's simply $\langle(1-y)\rangle=\int_0^1 (1-y)dy=1/2$ hence the
result follows.

Let us now turn to the study of the statistical properties of adaptive
walks. The two major outcomes are concerned
with the average length of an adaptive walk, which represents the average
number of configurations the system has assumed from its starting one to a
local fitness maximum, and with the average
duration of an adaptive walk, namely the
total number of tried mutations, those accepted and those refused.
\begin{description}
\item[Result 3.] Let $\langle\ell(N)\rangle$ and $\langle t(N)\rangle$
denote, respectively, the average length and the average duration of an
adaptive walk. If $N\gg 1$ we have
\begin{enumerate}
\item $\langle\ell(N)\rangle\simeq\log_2 N$ ;
\item $\langle t(N)\rangle\simeq N$.
\end{enumerate}
\end{description}
1. is an estimate for $\langle\ell(N)\rangle$. It is obtained through the
consideration that an adaptive walk ends when the fraction $F(\ell)$ of fitter
neighbours falls below $1/N$. Hence the average length is that for which 
$\langle F(\ell)\rangle\simeq\frac{1}{N}$. From Result 2 one soon gets
\begin{equation}
2^{-\ell}\simeq\frac{1}{N} ,
\end{equation}
whence the estimate $\langle\ell(N)\rangle\simeq\log_2 N$ follows.
A more rigorous though much more complicated estimate has been derived in
Ref. \cite{flyvbjerg-lautrup}, where it is shown that 
$\langle\ell(N)\rangle\simeq\log N$ with a
proportionality constant that is slightly different from $(\log 2)^{-1}$.
Therefore it is reasonable to take 1. as a fairly good estimate.

For what concerns 2., we consider that since the fraction of
fitter neighbours is halved on the average at each step, then the waiting
time (if one wants, the number of tried and refused mutations) doubles
on the average at each step. So the average number of time units one
has to wait in order to take the $\ell$-th step is $2^{\ell-1}$ (one has to
wait a time $1$ to take the first step because by assumption each walk
starts from the least fit configuration in $\Gamma$). We obtain
$\langle t(N)\rangle$
by summing all waiting times in each configuration passed by in an adaptive
walk, the average number of which is given by 1.; hence
\begin{equation}
\langle t(N)\rangle=\sum_{\ell=1}^{\log_2 N}2^{\ell-1}=N-1,
\end{equation}
where the sum has been performed as if $\log_2 N$ were an integer.
For large $N$ 2. is recovered. Again, in Ref. \cite{flyvbjerg-lautrup}
it is shown that a more rigorous derivation of $\langle t(N)\rangle$ yields
the same result $\langle t(N)\rangle\simeq N$ up to a proportionality
constant that is just slightly different from $1$. Therefore 2. may well
be considered a good estimate.

\section{Perturbing Kauffman's model}
In the previous section we have recalled the statistical properties of
Kauffman's maximally rugged model.
Following its dynamical rule $F$ the system can evolve only through fitter
configurations.
In some sense, looking back at spin glasses, one could say that it
lacks of frustration. The system always does the right thing,
always finds its way in the rugged landscape, in a finite number of steps
reaches a fitness maximum, and that's it; failures are ruled out.
In our perturbed version
of this model we want to frustrate the rationality of the system 
with an additive selective pressure $p$, acting as a constraint on the
system's optimizing ability.

We thus consider a system whose phase space is $\Gamma=\{+1,-1\}^N$, evolving
in a landscape where the fitness $\phi$ is a quenched random variable.
The law of evolution $F_p$ depends on a real parameter $p\in[0,1]$ through
the following definition:
\begin{enumerate}
\item at time $t\geq 0$ the system is in configuration $\mathcal{C}_t=\{S_1,
\ldots,S_N\}$ with fitness $\phi(\mathcal{C}_t)$;
\item a spin $S_i$ of $\mathcal{C}_t$ is chosen at random and its sign is
changed, thus obtaining a configuration $\mathcal{C'}$ that differs from
$\mathcal{C}_t$ by just the $i$-th spin ($1\leq i\leq N$);
\item if $\phi(\mathcal{C'})>\phi(\mathcal{C}_t)$ then, with
probability $p$, $\mathcal{C}_{t+1}=\mathcal{C'}$ and, with probability
$1-p$, $\mathcal{C}_{t+1}$ is chosen at random among the $N$ neighbouring
configurations of $\mathcal{C}_t$;
\item if $\phi(\mathcal{C'})<\phi(\mathcal{C}_{t})$, then
$\mathcal{C}_{t+1}=\mathcal{C}_t$ and return to (ii).
\end{enumerate}

The landscape's statistical properties are the same as those of
Kauffman's model, so that Result 1 still holds.
The difference with the original model is that this time the system
accepts a favourable mutation only with a probability $p$. If it
can not, then it is forced to choose a random spin and change its sign,
regardless of the fitness of this newly-obtained configuration. By this
we mean to model a system that undergoes an external
evolutionary pressure, whose strength increases with $p$ varying from
$1$ to $0$, as it evolves in a rugged landscape. The pressure is
a perturbation of the dynamics, such that the case $p=1$ corresponds to
the unperturbed model. We'll see that a small perturbation is sufficient
to drastically change all statistical properties of the model. For
example, the average length of a trajectory, which we shall call
a \emph{p-walk}, diverges.

Let us consider the case of finite $N$. We begin by deriving the analogous
for the perturbed model of Result 2 for the unperturbed one.
\begin{description}
\item[Result 4.] Let $F_p(\ell)$ denote the fraction of fitter neighbours
after $\ell$ steps of a $p$-walk and let $\langle F_p(\ell)\rangle\equiv
\langle F(\ell)\rangle_p$. We have
\begin{equation}
\langle F(\ell)\rangle_p=\frac{1}{2-p}\bigg(\Big(
\frac{p}{2}\Big)^{\ell}+1-p\bigg).
\end{equation}
\end{description}

One sees that in the $p\rightarrow 1$ limit, Result 2 is recovered.

The proof is not difficult but tedious.
Observe that at each mutation the system makes a choice between two
symbols: $p$ and $1-p$. Let $\Omega$ denote the set of all possible sequences
of choices the system can make in a $p$-walk of given length $\ell$,
namely $\Omega=\{p,1-p\}^\ell$. One can
think of a $p$-walk $\omega$ of $\ell$ steps as an element of $\Omega$ of the
form
\begin{equation}
\omega=\{\omega_1,\ldots,\omega_\ell\} ,
\end{equation}
where $\omega_j\in\{p,1-p\}$ for $1\leq j\leq\ell$. We shall call $\Omega$
the ``space of $p$-walks''.
Considering that when the system accepts a positive mutation the average
fraction of fitter neighbours is halved, one can construct a partition
of $\Omega$ made by subsets of ``similar'' $p$-walks:
\begin{enumerate}
\item the first subset $\Omega_{1}$ contains those $p$-walks such that
$\omega_{\ell}=1-p$;
\item the second subset $\Omega_{2}$ contains those $p$-walks such that
$\omega_{\ell -1}=1-p$ and $\omega_{\ell}=p$;
\item the $k$-th subset ($1\leq k\leq\ell -1$) $\Omega_{k}$
contains those $p$-walks such that
$\omega_{\ell -k+1}=1-p$ and $\omega_{\ell -k+2},\ldots,\omega_{\ell}=p$;
\item the $\ell$-th subset $\Omega_{\ell}$ contains the
$p$-walks $\{1-p,p,\ldots,p\}$ e $\{p,\ldots,p\}$.
\end{enumerate}
We shall call ``types'' of $p$-walks the subsets $\Omega_m$ ($m=1,\ldots,\ell$),
so that a $p$-walk $\omega\in\Omega_m$ is a $p$-walk of the $m$-th
type. The similarity consists of the fact that all $p$-walks of the $m$-th
type are such that, on the average, after the $\ell$-th
step there is a fraction of $2^{-m}$ fitter neighbours than the configuration
reached by the system. This is so because this average fraction depends
on how many $p$-steps (steps in which the mutation has been accepted) the
system has made since the last $(1-p)$-step. In fact, a $(1-p)$-step
brings the system to a configuration having, on the average, a fraction
of $1/2$ fitter neighbours (if $N$ is sufficiently large) and each
$p$-step following halves this fraction. For example, if $n-1$ $p$-steps
are taken after a $(1-p)$-step (namely if a $p$-walk of the
$n$-th subset is made), the average fraction of fitter neighbours
will be $2^{-n}$. So the probability to have $2^{-m}\cdot N$ fitter
neighbours
after $\ell$ steps equals the probability $P(\Omega_m)$
to take a $p$-walk of the $m$-th type. This is easily calculated:
the probability $P(\omega)$ that the $p$-walk $\omega=\{\omega_1,\ldots,
\omega_\ell\}$ is made is simply
\begin{equation}
P(\omega)=\prod_{i=1}^{\ell}\omega_i 
\end{equation}
and thus
\begin{equation}
P(\Omega_m)=\sum_{\omega\in\Omega_m}P(\omega) .
\end{equation}
A straightforward calculation shows that
\begin{align} \label{pomegam}
P(\Omega_\ell) &= p^{\ell -1}\\
P(\Omega_k) &= (1-p)p^{k-1}
\end{align}
with $1\leq k\leq\ell -1$. Hence $\langle F(\ell)\rangle_p$ may be derived
from the formula
\begin{equation} \label{formula}
\langle F(\ell)\rangle_p=\sum_{m=1}^{\ell}P(\Omega_m)2^{-m} ,
\end{equation}
that makes use of the fact that the average fraction of fitter neighbours is
$2^{-m}$ with probability $P(\Omega_m)$ (namely, when the walk done is of
the $m$-th type), and of the fact that, clearly,
\begin{equation}
\sum_{m=1}^{\ell}P(\Omega_m)=1 .
\end{equation}
We rewrite formula (\ref{formula}) explicitly:
\begin{equation}
\langle F(\ell)\rangle_p=p^{\ell -1}2^{-\ell}+\sum_{k=0}^{\ell-2}(1-p)p^{k}2^{-(k+1)}.
\end{equation}
Performing the sum and with a minor rearrangement of the terms Result 4 is
obtained.

Result 4 is the starting point for deriving an estimate for the average
length $\langle\ell(N)\rangle_p$ of a $p$-walk.
It is sufficient to consider
that on the average the walk stops when $\langle F(\ell)\rangle_p$ falls below
the value $\frac{1}{N}$, that is, when there are no fitter neighbours.
Hence the stopping condition reads
\begin{equation}
\langle F(\ell)\rangle_p\sim\frac{1}{N} ,
\end{equation}
which leads to
\begin{equation}
\Big(\frac{p}{2}\Big)^{\ell}+1-p\sim\frac{2-p}{N} .
\end{equation}
Isolating $\ell$ from the previous formula is a
simple task and one obtains
\begin{equation} \label{ellemedia}
\langle\ell(N)\rangle_p\simeq\log_{\frac{p}{2}}\Big[\frac{1}{N}
\big((N-1)p-(N-2)\big)\Big] .
\end{equation}
One sees that in the $p\rightarrow 1$ limit the average length of an
adaptive walk is recovered:
\begin{equation}
\langle\ell(N)\rangle_1\equiv\langle\ell(N)\rangle\simeq\log_2 N .
\end{equation}

Formula (\ref{ellemedia}) may be put in a more fashionable way as is shown
by the following result, analogous to Result 3 of the unperturbed model.
\begin{description}
\item[Result 5.] Let $\langle\ell(N)\rangle_p$ and $\langle t(N)\rangle_p$
denote, respectively, the average length and duration of a $p$-walk.
There exists an $N$-dependent number $p_c\in]0,1[$ such that the following
estimates hold:
\begin{enumerate}
\item
\begin{equation} \label{ellemediaconpc}
\langle\ell(N)\rangle_p\simeq\frac{\log\Big[\frac{1}{N}
\big((p-p_c(N))(N-1)\big)\Big]}{\log\frac{p}{2}} ;
\end{equation}
\item
\begin{equation}
\langle t(N)\rangle_p\simeq\frac{1}{1-2p}\bigg((1-p)
\langle\ell(N)\rangle_p-p\frac{1-(2p)^{\langle\ell(N)\rangle_p}}
{1-2p}\bigg).
\end{equation}
\end{enumerate}
\end{description}
Let us rewrite formula (\ref{ellemedia}) in the form
\begin{equation}
\langle\ell(N)\rangle_p\simeq\frac{\log\Big[\frac{1}{N}
\big((N-1)p-(N-2)\big)\Big]}{\log\frac{p}{2}} .
\end{equation}
This must by definition be a positive number, though it may not be an
integer. But since its denominator is negative, so has to be its numerator.
But this only holds if
\begin{equation} 
0<\frac{1}{N}\big((N-1)p-(N-2)\big)<1.
\end{equation}
The right side inequality leads to a condition for $p$ that is always
satisfied; the left inequality leads on the contrary to the requirement
that
\begin{equation} \label{pici}
p>p_c(N)=1-\frac{1}{N-1}.
\end{equation}
Minor rearrangements of the terms in formula (\ref{ellemedia}) lead thus
to part 1. of Result 5.

We see now from formula (\ref{ellemediaconpc}) that the average length of a
$p$-walk diverges as $p\rightarrow p_c(N)^+$. Note that the critical
threshold $p_c$ depends on the dimension $N$ of the phase space. Also,
it is simple to check that, for all $N$,
\begin{equation}
\lim_{p\rightarrow p_c^+}\frac{\langle\ell(N)\rangle_p}{\log(p-p_c)}=\beta<0.
\end{equation}
This means that when we approach the critical point from above the average
length diverges as
\begin{equation}
\langle\ell(N)\rangle_p\approx\log(p-p_c)^\beta=-|\beta|\log(p-p_c).
\end{equation} 

Let us turn to time. It is clear that a $p$-walk of infinite
length is also of infinite duration. Reminding that one unit of time
corresponds to a trial spin flip and fitness check in our model,
let us derive an expression for the average duration $\langle t(N)\rangle_p$
of a $p$-walk. The strategy we wish to adopt is the following: since the
system remains for a certain amount of time in each configuration it visits,
during which it tries point mutations to find a fitter neighbour,
we can think of evaluating the average time spent in a configuration, and
sum over all configurations visited during a $p$-walk, that on the average
are $\langle\ell(N)\rangle_p$. In the case of an adaptive walk everything's
simple: since the fraction of fitter neighbours is halved on the average
at each step, the waiting time is doubled on the average, so that
in order to take the $\ell$-th step it is necessary to wait a time
$\langle\tau(\ell)\rangle=2^{\ell-1}$
on the average (the average waiting time to take the first step is one
since all neighbours are fitter by assumption, and so on). 

Hence, when the average fraction of fitter neighbours is $2^{-\ell}$ the
time required on the average to find a fitter one mutant configuration is
$2^{\ell-1}$. We thus could estimate the average waiting time
$\langle\tau(\ell)\rangle_p$
to take the $\ell$-th step of a $p$-walk as we
did estimate $\langle F(\ell)\rangle_p$, namely by formula (\ref{formula}),
just by substituting
all average fractions of fitter neighbours $2^{-\ell},\ldots,2^{-1}$
with the corresponding average waiting times
$2^{\ell -1},\ldots,1$. We have
\begin{equation}
\langle\tau(\ell)\rangle_{p}=\sum_{m=1}^{\ell} P(\Omega_{m})2^{m-1} ;
\end{equation}
using the probabilities (\ref{pomegam}) we obtain immediately
\begin{equation}
\langle\tau(\ell)\rangle_p=p^{\ell -1}2^{\ell -1}+(1-p)
\sum_{k=0}^{\ell -2}p^{k}2^{k}.
\end{equation}
Performing the sum we finally arrive at
\begin{equation}  \label{taumedio}
\langle\tau(\ell)\rangle_p=\frac{1}{1-2p}\Big(1-p-p(2p)^{\ell -1}\Big) ,
\end{equation}
which is what we were looking for. One sees that in the no-perturbation
limit $p\rightarrow 1$ this result leads to
$\langle\tau(\ell)\rangle_1=2^{\ell -1}$, as we expected. Furthermore, note
that for $\ell=1$ we get $\langle\tau(1)\rangle_p=1$ for every $p$, which is
right since in this model also the time needed to take the first step
is one.

Summing over all configurations transversed during a $p$-walk on the average
we obtain an estimate for the average duration $\langle t(N)\rangle_p$:
\begin{equation}
\langle t(N)\rangle_p\simeq\sum_{\ell=1}^{\langle\ell(N)\rangle_p}
\langle\tau(\ell)\rangle_p=
\sum_{\ell=0}^{\langle\ell(N)\rangle_p-1}\langle\tau(\ell)\rangle_p .
\end{equation}
The last equality holds by virtue of the fact that in formula
(\ref{taumedio}) the dependence on $\ell$ is only in the term
$(2p)^{\ell -1}$. Therefore we may redefine $\ell$ (ranging from
$1$ to $\langle\ell(N)\rangle_p$) as $\ell -1$, and this new
variable varies from $0$ to $\langle\ell(N)\rangle_p -1$.
After a minor rearrangement of the terms we see we can split the sum in
two sums:
\begin{equation}
\langle t(N)\rangle_p\simeq\frac{1}{1-2p}\Bigg(\sum_{\ell =0}^{\langle
\ell(N)\rangle_p -1}(1-p)-
p\sum_{\ell =0}^{\langle\ell(N)\rangle_p -1}(2p)^{\ell}\Bigg).
\end{equation}
Now perform the sums under the assumption that $\langle\ell(N)\rangle_p$
is an integer (we are
interested in an estimate; if the average length were not an integer,
we would get an estimate by summing up to $\lfloor\langle\ell(N)
\rangle_p\rfloor$):
\begin{equation}
\langle t(N)\rangle_p\simeq\frac{1}{1-2p}\bigg((1-p)\langle\ell(N)\rangle_p -
p\frac{1-(2p)^{\langle\ell(N)\rangle_p}}{1-2p}\bigg),
\end{equation}
which is the estimated average duration of a $p$-walk. Note that in the
$p\rightarrow 1$ limit, where $\langle\ell(N)\rangle_p\rightarrow\log_2 N$,
one recovers
\begin{equation}
\langle t(N)\rangle_1\equiv\langle t(N)\rangle\simeq N-1 ,
\end{equation}
that in the large $N$ limit is what one gets from Kauffman's maximally
rugged model.

We have thus discovered that if $p>p_c(N)$ then the average length of a
$p$-walk in the rugged random landscape is finite, whereas 
at $p=p_c(N)$ $\langle\ell(N)\rangle_p$ diverges. We emphasize that this
picture is qualitatively correct, despite of the fact that formula
(\ref{ellemediaconpc}) is an estimate for $\langle\ell(N)\rangle$.
The critical probability
$p_c(N)$ depends on the phase space's dimension $N$. If $N$ is large,
as we have assumed to derive these formulas, then it is close to one.
Thus a small perturbation, which means a value of $p$ which is just slightly
different from $1$, is sufficient to switch
on the probability that the system wanders through the rugged
landscape indefinitely. 

In effect, we can render these observations
more quantitative.
\begin{description}
\item[Result 6.] Let $Q_p(\ell)$ denote the probability that a
$p$-walk consists of $\ell$ steps. We have
\begin{equation} \label{cases}
Q_p(\infty)
\begin{cases}
=0& \text{for $p>p_c(N)$};\\
>0& \text{for $p<p_c(N)$}.
\end{cases}
\end{equation}
\end{description}
Of course we assume the validity of the normalization condition
\begin{equation}
Q_p(\infty)+\sum_{\ell=1}^{\infty}Q_p(\ell)=1,
\end{equation}
that should hold for all $p\in[0,1]$ and where we have separated the term
corresponding to $\ell=\infty$.
To prove Result 6 it is sufficient to put $\langle\ell(N)\rangle_p$ in the
form
\begin{equation} \label{nonstandard}
\langle\ell(N)\rangle_p=\infty Q_p(\infty)+\sum_{\ell=1}^{\infty}\ell Q_p(\ell) ,
\end{equation}
and consider that this average value is finite whenever $p>p_c(N)$, and
infinite otherwise.

We have mutuated this fancy way of writing
this average value from percolation theory, where the average
number of lattice points in a cluster $\langle n\rangle_p$ is written as
(see for example Ref. \cite{grimmett}, where $\langle n\rangle_p\equiv
\chi(p)$)
\begin{equation} \label{nonstandard2}
\langle n\rangle_p=\infty P_p(\infty)+\sum_{n=1}^{\infty}n P_p(n) ,
\end{equation}
where $P_p(n)$ represents the probability that a cluster contains
exactly $n$ points.
Equations (\ref{nonstandard}) and (\ref{nonstandard2}) are not very
satisfactory from a notation point of view, since the quantity $\infty$
is treated like a number. Anyhow, equation (\ref{cases}) indicates
that this $p$-walks' model displays a percolation-like transistion, that
is just analogous to the one described by
\begin{equation}
P_p(\infty)
\begin{cases}
=0& \text{for $p<p_c(N)$};\\
>0& \text{for $p>p_c(N)$},
\end{cases}
\end{equation}
which characterises ``classical'' percolation theory,
where $N$ indicates the dimensionality of the lattice one is considering.
In the large $N$ limit of the $p$-walks' model the percolation threshold
is close to $1$ (see formula (\ref{pici})), so that a small perturbation is
enough to turn on the probability of no arrest.

\section{The thermodynamic limit}
The thermodynamic limit is obtained by letting the dimension $N$ of
the phase space $\Gamma=\{+1,-1\}^N$ to infinity. In this limit the average
length of an adaptive walk diverges logarithmically as stated by Result 3.
Hence, all $p$-walks are interminable. It is therefore natural to consider,
together with the $N\rightarrow\infty$ limit for $\Gamma$, the
$\ell\rightarrow\infty$ limit for $\Omega=\{p,1-p\}^\ell$.

In the previous section we have seen that for finite $\ell$ $\Omega$ may be
fragmented into $\ell$
subsets which we called ``types'' of $p$-walks. All walks of the same type,
say the $m$-th ($1\leq m\leq\ell$), are such that, on the average, after
$\ell$ steps the fraction of fitter neighbours or, if one wants, the
probability to take one further step is $2^{-m}$. We have denoted by
$\Omega_m$ the $m$-th type and by $P(\Omega_m)$ the probability that
a $p$-walk is of the $m$-th type. We have then found that
$P(\Omega_k)=(1-p)p^{k-1}$ for $k=1,\ldots,\ell -1$, and that
$P(\Omega_\ell)=p^{\ell -1}$. When $\ell$ goes to infinity the number of
fragments in which the $\Omega$ space is broken diverges, but still
each fragment retains the same meaning, for the probability that a
certain $p$-walk is of a given type does not change if the number of types
diverges. For example, the probability that a $p$-walker finds half
of his neighbours fitter than him after $\ell$ steps is always $1-p$ for
all $\ell$. What happens is just that when the walker takes an $(\ell +1)$-th
step an additional type (the $(\ell +1)$-th) must be taken into account.
But its probability $P(\Omega_{\ell +1})$ causes a change in $P(\Omega_\ell)$,
whereas the probabilities of the remaining types are unchanged.

Hence, $\Omega$'s thermodynamic limit may be thought of as if it were
constructed as follows. Take $\Omega$ as an object of size $P(\Omega)=1$
and suppose to break it into infinite pieces $\Omega_1,\Omega_2,\ldots$
of sizes $W_1=P(\Omega_1),W_2=P(\Omega_2),\ldots$ respectively.
The breaking process depends on a given real number $p\in[0,1]$.
First, we tear $\Omega$ in two pieces of sizes $W_1=1-p$ and $p$.
Then we take the latter and tear it in two pieces of sizes
$W_2=(1-p)p$ and $p^2$. Thirdly, we take the one of size $p^2$ and
break it in two pieces of sizes $W_3=(1-p)p^2$ and $p^3$. In principle,
one may continue breaking the pieces of sizes $p^{\ell +1}$ at the $\ell$-th
step and take the $\ell\rightarrow\infty$ limit. In the end we have
an infinite set of pieces of sizes
\begin{align}
W_1 &= 1-p  \nonumber\\
W_2 &= (1-p)p \nonumber\\
\ldots & \\
W_s &= (1-p)p^{s-1} \nonumber\\
\ldots & \nonumber 
\end{align}
The sizes $W_s$ represent the probability that a $p$-walk is of the
$s$-th type in the thermodynamic limit. Clearly, $\sum_s W_s=
\sum_{s=1}^{\infty}(1-p)p^{s-1}=1$. In Ref. \cite{adm} we called this a
\emph{geometrically broken object}, since the sizes of the resulting pieces
form a geometric sequence. In fact, $W_{s+1}=pW_s$ for $s=1,2,\ldots$ .

Now suppose to be given a certain number of $p$-walkers, each of which
chooses his value of $p$ from a given probability density $\rho(p)$ on
the $[0,1]$ interval. To each of these will correspond a specific rupture
of the space $\Omega$ of $p$-walks, since for the geometrical breaking
the weights of the types $W_s$ depend just on the value of $p$. Hence
each $p$-walker gives a breaking sample of $\Omega$. This picture is
quite usual in the theory of disordered systems, where one deals with
systems having a quenched disorder represented by a number of stationary
random variables. For each sample, namely for each choice of the quenched
disorder, certain statistical or thermodynamic extensive obsevables $X$
(for example, the free energy density) may be evaluated.
One is usually interested in averaging $X$ over disorder, i.e. over all
possible samplings of the quenched random variables. The most
interesting outcome in many cases is that non self-averaging effects
are present: sample-to-sample fluctuations of $X$ do
not vanish in the thermodynamic limit (i.e. when one lets the size of
the sistem go to infinity). This means that $\langle X\rangle$
(the average of $X$ over disorder) is finite and that
$\textrm{var}(X)=\langle X^2\rangle-\langle X\rangle^2$ is non zero.
The probability density $P(X)$ of $X$ remains ``broad'' in the
thermodynamic limit, whereas for a self-averaging quantity the
probability density in the same limit is highly concentrated around its
average. As a result, the value of a self-averaging quantity on a
sufficiently large sample is a good estimate of the ensemble average,
while for non self-averaging quantities no sample, no matter how large,
is a good representative of the whole ensemble.

More specifically, in model broken
objects as the randomly broken object \cite{rbo} as well as in other more
complicated models (see Ref. \cite{overview} for a unifying review) one
finds that the sizes $W_s$ of the pieces lack of self-averaging.
In all of these the thermodynamic limit is obtained by letting the
number of pieces go to infinity.
The study of non self-averaging properties of a geometrically broken object
is the content of Ref. \cite{adm}. The model turns out to be exactly solvable.

We consider in each sample the probability
\begin{equation}
Y=\sum_s W_s^2
\end{equation}
that two randomly chosen walks in $\Omega$ are of the same type. The aim is
to show that $Y$'s ensemble average $\langle Y\rangle$ over disorder
(that is, over $p$) is non zero and that $Y$'s variance $\textrm{var}(Y)=
\langle Y^2\rangle -\langle Y\rangle^2$ does not vanish. This would
yield the conclusion that the probabilities $W_s$ of the types are non
self-averaging quantities. Among other results, in Ref. \cite{adm} we have
proved that
\begin{enumerate}
\item the probability density $\Pi(Y)$ of $Y$ over all possible samples
of a geometrically broken object is given in the thermodynamic limit by
\begin{equation}
\Pi(Y)=\frac{2}{(1+Y)^2}\rho\bigg(\frac{1-Y}{1+Y}\bigg).
\end{equation}
\item Assuming $\rho(p)=1$ the ensemble average of $Y$ is given by
\begin{equation}
\langle Y\rangle=\int_0^1Y\Pi(Y)dY=\log 4 -1\simeq 0.386\ldots.
\end{equation}
\item Under the same assumption one can calculate the second moment
$\langle Y^2\rangle$ of $Y$ and show that the variance is given by
\begin{equation}
\textrm{var}(Y)=\langle Y^2\rangle-\langle Y\rangle^2\simeq 0.078\ldots.
\end{equation}
\end{enumerate}

We thus come to the interesting conclusion that in the thermodynamic limit
of the $p$-walks' model non self-averaging effects are present: the
probabilities that a $p$-walk is of a given type lack of self-averaging
(i.e. they remain sample dependent). In other terms, the probability
$Y$ that two $p$-walkers with same $p$ make walks of the same type has
non zero average and finite variance, despite of the fact that there
are infinite different types.

Let us now turn to a different problem. Consider two $p$-walkers with
freedom parameters $p_1$ and $p_2$ respectively. We know that for each of
them the probability that a $p$-walk is of the $s$-th type is given by
$W_s(p_i)=(1-p_i)p_i^{s-1}$, for $i=1,2$. Let us define the variable
\begin{equation}
Z\equiv Z(p_1,p_2)=\sum_s W_s(p_1)W_s(p_2),
\end{equation}
giving the probability that a randomly chosen $p_1$-walk and a randomly
chosen $p_2$-walk in $\Omega$ are of the same type. The ensemble average
$\langle Z\rangle$
has to be evaluated over all possible choices of $p_1$ and $p_2$.
$\langle Z\rangle$ has some resemblence with a correlation function in the
space $\Omega$ of $p$-walks. We shall now prove that it is possible to
calculate the probability density $\Phi(Z)$ of $Z$, such that the probability
that for a given choice of $p_1$ and $p_2$ $Z$ is
in the $[Z,Z+dZ]$ interval is given by $\Phi(Z)dZ$. More precisely we
prove the following:
\begin{description}
\item[Result 7.] If both $p_1$ and $p_2$ are chosen from a uniform probability
density on the $[0,1]$ interval, then the probability density $\Phi(Z)$
of $Z$ is given by
\begin{equation}
\Phi(Z)=\frac{2}{(1+Z)^3}\big(1-Z^2-2Z\log Z\big).
\end{equation}
\end{description}

This allows us to evaluate $\langle Z\rangle$ and $\textrm{var}(Z)=
\langle Z^2\rangle-\langle Z\rangle^2$. One finds that
\begin{align}
\langle Z\rangle &= \int_0^1 Z\Phi(Z)dZ\simeq 0.289867\ldots\\
\langle Z^2\rangle &= \int_0^1 Z^2\Phi(Z)dZ\simeq 0.130395\ldots\\
\textrm{var}(Z) &\simeq 0.04637\ldots
\end{align}
This tells us that, like $Y$, $Z$ is non self-averaging. But it also tells
us that the values of $Z$ are more concentrated around its average than
those of $Y$, at least for uniform $\rho$, since $\textrm{var}(Z)$ is smaller
than $\textrm{var}(Y)$.

\begin{figure}
\begin{center}
\epsfig{file=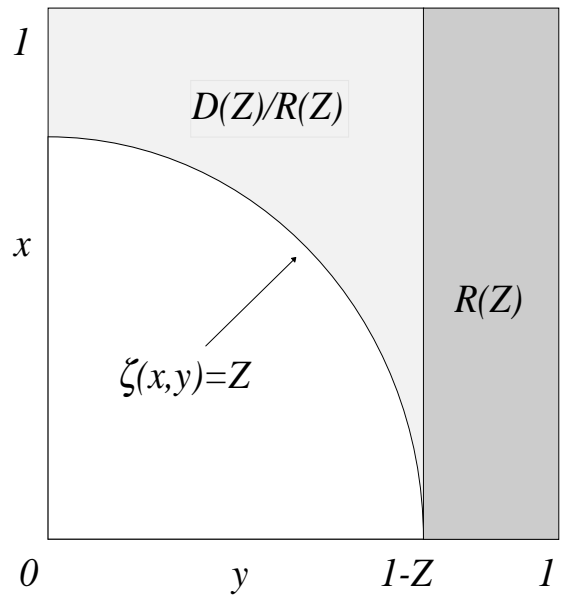,height=12cm,width=8cm}
\caption{Regions $R(Z)$ and $D(Z)\setminus R(Z)$.}
\label{fig:fig}
\end{center}
\end{figure}

We begin by calculating $Z$ for two given values of $p_1$ and $p_2$.
One has
\begin{equation}
Z=\sum_{s=1}^{\infty}(1-p_1)p_1^{s-1}(1-p_2)p_2^{s-1}=
(1-p_1)(1-p_2)\sum_s(p_1 p_2)^{s-1},
\end{equation}
whence
\begin{equation}
Z(p_1,p_2)=\frac{(1-p_1)(1-p_2)}{1-p_1 p_2} .
\end{equation}
For simplicity of notation set $p_1=x$ and $p_2=y$. Let
\begin{equation} \label{zeta}
\zeta(x,y)=\frac{(1-x)(1-y)}{1-xy}
\end{equation}
and define the region $D(Z)\subseteq[0,1]^2$
\begin{equation}
D(Z)=\{(x,y)\in[0,1]^2 : (\zeta(x,y)\leq Z) \wedge (Z\in[0,1])\}.
\end{equation}
Suppose that $x$ and $y$ are random variables with probability distributions
$\rho(x)$ and $\rho(y)$. The probability $\mathsf{P}\{\zeta(x,y)\leq Z\}
\equiv F(Z)$ is simply
\begin{equation}
F(Z)=\int\!\!\!\int_{D(Z)}\rho(x)\rho(y)dxdy.
\end{equation}
$F(Z)$ and $\Phi(Z)$ are related by
\begin{equation}
\Phi(Z)=\frac{dF(Z)}{dZ}
\end{equation}
therefore calculating $F(Z)$ is the crucial step towards $\Phi(Z)$.
Suppose for simplicity that $\rho=1$, so that $F(Z)$ is the area
of $D(Z)$.
From definition (\ref{zeta}) we see that for $x=0$ $\zeta(0,y)=1-y$.
Hence the curve $\zeta(x,y)=Z$ touches the $y$ axis in the point
$y_0=1-Z$. We thus construct the rectangle $R(Z)=[y_0,1]\times[0,1]$
(as shown in Figure \ref{fig:fig})
and note that it is contained in $D(Z)$. $F(Z)$ may thus be
separated as
\begin{equation} \label{effedizeta}
F(Z)=\int\!\!\!\int_{R(Z)}dxdy+\int\!\!\!\int_{D(Z)\setminus R(Z)}dxdy.
\end{equation}
The first integral is equal to the area of $R(Z)$, that is $Z$.
For what concerns the second integral, we choose to evaluate it for
$x$ running on the curve $\zeta(x,y)=Z$ and $y$ ranging from $0$ to
$y_0=1-Z$. The coordinates $x$ of the points on the curve
$\zeta(x,y)=Z$ have the form
\begin{equation}
x(y,Z)=\frac{y+Z-1}{y(Z+1)-1},
\end{equation}
as can be seen by inversion of definition (\ref{zeta}). Therefore
\begin{equation}
\int\!\!\!\int_{D(Z)\setminus R(Z)}dxdy=\int_0^{1-Z}dy\int_{x(y,Z)}^1 dx=
\int_0^{1-Z}\big(1-x(y,Z)\big)dy.
\end{equation}
\nopagebreak
We thus need to calculate the integral
\begin{equation}
\int\!\!\!\int_{D(Z)\setminus R(Z)}dxdy=\int_0^{1-Z}\Bigg(
1-\frac{y+Z-1}{y(Z+1)-1}\Bigg)dy .
\end{equation}
This is a quite simple task, and the result is
\begin{equation}
\int\!\!\!\int_{D(Z)\setminus R(Z)}dxdy=
\frac{Z}{(1+Z)^2}\big(1-Z^2-2Z\log Z\big).
\end{equation}
We finally obtain $F(Z)$ from identity (\ref{effedizeta}):
\begin{equation}
F(Z)=\frac{2Z}{(1+Z)^2}\big(1+Z(1-\log Z)\big).
\end{equation}
Differentiating this with respect to $Z$ we get at last Result 7:
\begin{equation}
\Phi(Z)=\frac{2}{(1+Z)^3}\big(1-Z^2-2Z\log Z\big).
\end{equation}

\section{Conclusion}
To summarize we have studied an abstract evolutionary model in which
the system's size is $N$ and phase space $\Gamma$ has $2^N$ configurations.
The evolutionary rule $F_p$ is a stochastic map that depends on a
real parameter $p\in[0,1]$. For $p=1$ we recover Kauffman's maximally
rugged model and trajectories to local fitness optima are adaptive walks. For
generic $p$ we have introduced $p$-walks. In the finite $N$ case we have
shown that the average length of a $p$-walk as estimated by
Result 5 is finite whenever $p>p_c(N)$, where the critical value $p$ is
given by $p_c(N)=1-1/(N-1)$. When $p\rightarrow p_c(N)^+$ and for all
$p<p_c(N)$ the average length diverges. This results in a percolation-like
phase transition. In the supercritical phase ($p>p_c(N)$) all
$p$-walks are of finite length, whereas in the subcritical phase
($p<p_c(N)$) the probability of an infinitely long $p$-walk is non
zero. In the thermodynamic limit $N\rightarrow\infty$ we have emphasized
the fact that the dimension of the space $\Omega$ of $p$-walks must be
considered infinite. $\Omega$ contains all representations of $p$-walks of
a given length, hence the thermodynamic limit yields a divergence in the
number of different possible evolutions. We have shown that $\Omega$
may be partitioned in infinite subsets grouping ``similar'' $p$-walks.
This fragmentation is analogous to that of a geometrically broken
object. Hence, we were able to prove that non self-averaging
effects are present: the probability $Y$ that two $p$-walkers with same value
of $p$ have similar evolutions has non zero average and finite variance,
even though the number of different types of evolutions is infinite.
Lastly, we have studied the probability $Z$ that two different $p$-walkers
(with different values of $p$) have similar evolutions and have shown that
$Z$ is also non self-averaging. The simplicity of the model has made it
possible to obtain analytical results in the thermodynamic limit for
both $Y$ and $Z$.

These results deserve some comment.
The $p$-walks' model seems to be versatile for different metaphoric
interpretations, mostly because of its simple definition. Yet, it
has turned out to display a rich and non-trivial behaviour even in the
thermodynamic limit. It represents another non self-averaging model,
adding to a list which indicates the strong need to
find a more general theory, or at least the universality underlying
the presence of this phenomenon in many different contexts.
We have also stressed in the introduction
that we have worked out this model as a model of an abstract behaviour.
Nevertheless, a comparison with biological evolutionary models is
possible. Ref. \cite{higgs} offers a detailed account on the biological
side of non self-averaging effects. Interestingly, such quantities as
$Y$ in abstract disordered models are measurable quantities for
biological systems. More precisely in population genetics $Y$ corresponds
to a parameter called homozygosity, giving the probability that two
genes sampled randomly at the same locus in two individuals are
identical. It is an experimental
fact, as is explained in Ref. \cite{higgs}, that $Y$ has a broad distribution
for a large number of polymorphic loci in Drosophila. This can be a
convincing evidence of the fact that the evolutionary process is non
self-averaging. From this viewpoint, we think our model shows that
a less strict dynamical rule is necessary for non self-averaging effects
to appear in a Kauffman-type of model. If we are in a tightly adaptive
situation two systems undergoing biological evolution will always be
doing the same type of walk, which would mean $Y=1$ with a trivial
distribution. On the
contrary, if a certain variability is allowed, the probability $Y$
that the two systems find themselves in similar states is still not zero,
on the average, but its distribution is broad and non trivial. This
kind of evolution sounds to be closer to that implied by the
experimental results on Drosophila. Note that the existence of a variability
in the rule implies a non zero probability of failure, which in our model
is the very feature leading to non self-averaging effects.

\end{document}